\documentclass[pra,superscriptaddress,twocolumn]{revtex4-1}
\usepackage{amssymb,amsmath,amsthm}
\usepackage{color}
\usepackage{bm}
\usepackage{dsfont}
\usepackage{hyperref}
\usepackage{graphicx}
\usepackage{stackengine,scalerel}
\def\lambdabar{\ThisStyle{\ensurestackMath{\stackon[-2.4\LMpt]{%
  \SavedStyle\lambda}{\kern-.5pt\kern\LMpt\rule{1\LMex}{.25pt+.15\LMpt}}}}}

\definecolor{darkblue}{rgb}{0,0,0.5}
\definecolor{darkred}{rgb}{0.5,0,0}

\hypersetup{
  pdfborder={0,0,0},
  colorlinks=true,
  linkcolor=darkblue
}

\begin{document}

\title{Spontaneous emission and atomic line shift in causal perturbation theory}
\author{Karl-Peter Marzlin}
\affiliation{Department of Physics, St. Francis Xavier University,
  Antigonish, Nova Scotia, B2G 2W5, Canada}
\affiliation{Department of Physics and Atmospheric Science,
        Dalhousie University, Halifax, Nova Scotia B3H 4R2, Canada}
\author{Bryce Fitzgerald}
\affiliation{Department of Physics, St. Francis Xavier University,
 Antigonish, Nova Scotia, B2G 2W5, Canada}

\begin{abstract}
We derive spontaneous emission rate and line shift for two-level atoms
coupled to the radiation field using causal perturbation theory. In
this approach, employing the theory of distribution splitting prevents
the occurrence of divergent integrals. Our method confirms the result
for atomic decay rate but suggests that the cutoff frequency for the
atomic line shift is
determined by the atomic mass, rather than Bohr radius or electron mass.
\end{abstract}

\maketitle

\section{Introduction}
Quantum optics, the theory of the interaction between atoms and
photons, has been developed since the early days of the laser and has
led to many spectacular discoveries. It is an effective theory in
which atoms are reduced to point objects with just a few internal energy
eigenstates. Photons, 
described using the quantized electromagnetic field in Coulomb gauge,
trigger transitions between these states by inducing an electric
dipole moment. Hence, atoms are essentially modelled as a point
dipole for each transition between internal states.
When many-body effects are studied, each atomic energy level is
represented by a non-relativistic quantum field
\cite{PhysRevA.49.3799,PhysRevA.50.1681}, so that quantum optics
corresponds to a
non-relativistic quantum field theory. As it happens often in quantum
field theories, it is plagued by divergent integrals and
needs to be regularized. 

Like other effective theories, quantum optics possesses a natural smallest length scale:
the size of an atom. This length scale is often used as a cutoff to
regularize the theory, but this procedure is not universally applied.
Many researchers often simply ignore diverging
terms, stating that the line shift has been taken into account in the
definition of resonance frequencies. Others apply methods that are
similar to mass renormalization in high-energy physics. One may say
that there is no general agreement on which method should be used.

Despite the fact that different methods lead to different predictions,
this situation has not led to major problems. The reason is that very
often one is only interested in the decay rate of isolated atoms, which
does not depend on the regularization procedure. In dilute gases, the
variation of the Lamb shift with the thermal properties of an atomic
gas can also be safely ignored. However, this
non-rigorous approach may fail in situations where the
variation of line shifts with experimental parameters, such as
temperature, density, or distance to a dielectric material, becomes more pronounced.
Examples for such situations include spontaneous decay in photonic crystals
\cite{PhysRevA.50.1764,PRA59:2982} and absorbing dielectrics
\cite{PRA54:5227,PRA60:2534,dung:043816}, corrections to the
Lorentz-Lorenz formula in dense atomic gases \cite{PRA59:2427}, 
and plasmonics \cite{PhysRevA.85.063841}. It is therefore of
importance to find a more systematic approach. 

In this paper, we propose causal perturbation theory (CPT) as a
tool to make finite predictions for effective theories
such as quantum optics. We illustrate this approach by deriving 
spontaneous emission rate and line
shift for two-level atoms coupled to the radiation field.
After a short introduction to CPT in Sec.~\ref{sec:cpt},
we describe in Sec.~\ref{sec:cptsem} our strategy to derive decay rate and line shift.
Sec.~\ref{sec:T2} contains the derivation of the main result. The physical
implications of our results are discussed in Sec.~\ref{sec:discussion}.

\section{Causal perturbation theory}\label{sec:cpt}
The Wightman axioms \cite{StreaterWightman} specify that field
operators in quantum field theories correspond to operator-valued distributions.
For relativistic quantum field theories, Epstein and Glaser
\cite{EpsteinGlaser} have pointed out that ultraviolet divergencies
appear because of the improper splitting of
operator-valued distributions into retarded and
advanced part. They demonstrated that perturbation theory will
remain finite if one uses the proper theory of
distribution splitting \cite{Malgrange3} and causality. 
The details of how CPT is used in relativistic quantum electrodynamics
have been explained in Ref.~\cite{ScharfQED}, for instance.
We summarize the main aspects to explain the advantages of CPT for
effective theories.

Time evolution in quantum field theories can be described using 
a perturbation expansion of the S-matrix,
\begin{align} 
  S  &= \mathds{1} + \sum_{n=1}^\infty 
  \frac{ 1}{n!} \hat{T}_n
\\ 
   \hat{T}_n &=
\int d^4x_1 \cdots d^4 x_n\,
 T_n(x_1, \cdots , x_n)
  g(x_1)\cdots g(x_n),
\end{align} 
with $g(x)$ a test function that switches the interaction on and
off. 
We will denote four vectors by italic letters $x$
  and three-dimensional vectors by $\vec{ x}$. The components of
  these vectors are specified using greek and latin indices, respectively. 
The first-order term takes the form
\begin{align} 
  T_1(x) &= \frac{ -i}{\hbar c} H_\text{int}(x),
\end{align} 
where the interaction Hamiltonian $H_\text{int}(x)$ is typically a
product of field operators. In relativistic theories it is often more
convenient to express the first-order scattering amplitude in terms of
the Lagrangian. However, the starting point of non-relativistic
effective theories is usually the system Hamiltonian, so that we have
adopted this approach here.

Since $g(x)$ is a test function, integral 
$\int d^4x T_1(x) g(x)$ is well defined.
All higher order operators can be constructed
recursively. 
For our purpose, it suffices to consider the second-order term
\begin{align} 
  T_2(x,y) &= \left \{  \begin{array}{cc}
           A_2'(x,y) = -T_1(x) T_1(y)  & \text{for } x^0>y^0
    \\ 
           R_2'(x,y) = -T_1(y) T_1(x)  & \text{for } x^0<y^0
           \end{array} \right . .
\label{eq:T2def}\end{align} 
The usual procedure to separate $T_2(x,y)$ into its retarded and
advanced part would be to use step functions, 
\begin{align} 
  T_s(x,y) = -\theta(x^0-y^0) T_1(x) T_1(y) 
   -\theta(y^0-x^0) T_1(y) T_1(x). 
\label{eq:retardedNaive}\end{align} 
However, step functions are not test functions, so that this operation
is not well defined. Even though step functions may appear benign, they
are the cause for ultraviolet divergences. To illustrate this problem we
consider the well-known one-dimensional distribution 
$d(t) = (t\pm i \epsilon)^{-1}$, with $\epsilon\rightarrow 0$.
The Sokhotski-Plemelj theorem implies that
\begin{align} 
    \int_{-\infty}^{\infty} dt\, d(t) g(t)&=
     {\cal P} \int_{-\infty}^{\infty} dt\, \frac{ g(t)}{t}  
   \mp i \pi g(0),
\end{align} 
where ${\cal P}$ denotes the principal value.
If we instead consider the ``retarded'' distribution
$d_R(t) =\theta(t) d(t)$, we obtain
\begin{align} 
    \int_{-\infty}^{\infty} dt\, d_R(t) g(t) &=
    \lim_{\epsilon\rightarrow 0}\int_{\epsilon}^{\infty} dt\,
  \frac{ g(t)}{t}   \mp i \pi g(0).
\end{align} 
This integral is only well defined for test functions which fulfill
$g(0)=0$. In general, multiplying $d(t)$ with a step function will
therefore lead to divergent results.

The key difference between CPT and ordinary perturbation theory is
that Eq.~(\ref{eq:retardedNaive}) is avoided. Instead, retarded and
advanced part of a distribution are defined using distribution
splitting. Step function $\theta(t)$ is replaced by a
test function $\theta_L(t )$, which depends on a parameter $L$ and
converges toward $\theta(t)$ for $L\rightarrow 0$. For finite $L$,
the product $\theta_L(t) d(t)$ is then a well-defined distribution.
To avoid singularities in the limit $L\rightarrow 0$,
$g(t)$ is projected on the subspace of test functions for which the
limit is well-defined. This can be achieved by a kind of projection operator
$P$. In the example above, $P$ ensures that $g(0)=0$.
Distribution $\theta_L(t) d(t)P$ is then well-defined even in the limit
$L\rightarrow 0$. 

It is important to note that $P$ is generally not
unique. Different choices for $P$ can be characterized by a finite set of
parameters, which play a similar role as renormalization parameters in
conventional perturbation theory. In fact, it has been shown that CPT is
equivalent to the theory of renormalization for relativistic field
theories \cite{ScharfQED, ScharfYangMills}. One may say that CPT
provides an alternative approach to perturbation theory, which is more
rigorous with regard to distribution theory.

Since CPT is equivalent to renormalization theory and requires a
deeper understanding of distribution theory, it has only been used
by comparably few researchers to describe relativistic theories such as
QED \cite{GraciaAIP2006,AsteFiniteQFT2010} and
self-interacting quantum fields 
\cite{0305-4470-33-47-309, pinter2001finite,AsteCJP2003}.
It has also been applied to describe interacting quantum fields in
curved space-time \cite{FredenhagenJMP2016}.
In non-relativistic models, distribution splitting has been used 
to describe singular potentials in the Schr\"odinger equation 
\cite{10.3389/fphy.2014.00023}.

However, for effective theories CPT may help to clarify situations in which  
different methods produce competing results. Since the use of
perturbation theory with retarded and advanced Green's function is
very common, the clear identification of the origin of singular terms
in CPT can shed light on which methods are best suitable.
The main limitation
of CPT is that the underlying theory has to be causal. For effective
theories with relativistic fields, or for solid state systems with
finite propagation speed such as acoustic phonons,
this is already the case. Other effective theories, including quantum optics,
have to be modified to use CPT.
In the next section we demonstrate this procedure at the example of
spontaneous emission by a single two-level atom.

\section{Causal perturbation theory for spontaneous emission}\label{sec:cptsem}
 A cold gas of atoms interacting with light is well described by
representing the atoms by a set of non-relativistic field operators 
\begin{align} 
   \hat{\Psi}_n^{\text{(nr)}}(\vec{ x}) &= 
    (2\pi)^{-3/2} \int  d^3 k \, a^{(n)}_{\vec{k}} 
   e^{i \vec{ k}\cdot  \vec{ x}},
\end{align}    
which annihilate an atom with internal (electronic) energy $E_n$ at
position $\vec{ x}$
\cite{PhysRevA.49.3799,PhysRevA.50.1681}. Depending on the isotope,
atoms either correspond to Bosons or Fermions. In this paper, we will
consider bosonic atoms with two internal energy states, ground state
and excited state ($n=g,e$). 

CPT requires Einstein causality, so that we have to modify the
above model slightly by
replacing non-relativistic field operators by relativistic
fields.
In quantum optical models, the processes of exciting 
an atom,
\begin{align} 
   \hat{H}_\text{exc} &\sim \int d^3 x\,
    \hat{\Psi}_e^{\text{(nr)} \dagger  }(\vec{ x})
    \hat{\Psi}_g^{\text{(nr)}}(\vec{ x}),
\label{eq:QOexcHam}\end{align} 
and de-exciting the atom, $\hat{H}_\text{dex} = \hat{H}_\text{exc}
^\dagger $,  
are usually described by two separate terms in the Hamiltonian.
Since a real relativistic scalar field contains both annihilation and
creation operators for particles, it cannot be used to accomplish
such a construction. For this reason,
we represent atoms with internal
energy $E_n$ by a complex Klein-Gordon field
\begin{align} 
  \hat{\Psi}_n(x) &= (2\pi)^{-3/2} \int \frac{ d^3 k}{\sqrt{\lambdabar_n w_k}} 
  \left (
  a^{(n)}_{\vec{k}} e^{-ik\cdot x} + b^{(n)\dagger }_{\vec{k}} e^{ik\cdot x}
  \right ),
\label{eq:fieldOp}\end{align} 
where $k\cdot x = k_\mu x^\mu$
with $g_{\mu\nu}=$ diag(1,-1,-1,-1) and 
$k_\mu = (w_k,-\vec{ k})$, where
$ w_k=\sqrt{\vec{ k}^2 +
  \lambdabar_n^{-2} }$. The factor $\lambdabar_n = \hbar/(m_n c) $
corresponds to the Compton wavelength of the atom divided by $2\pi$.
The mass of an atom with internal energy $E_n$ is given by
$m_n = m_0 + E_n/c^2$. In our two-level model, the mass difference
between ground and excited state is related to the atomic resonance
frequency $\omega_{eg}$
via $(m_e-m_g)c^2 = \hbar\omega_{eg}$. This is in line with the
fact that the rest energy of a composite particle includes the binding
energy. 

Operators $a_{\vec{k}}^{(n)}, b_{\vec{k}}^{(n)}$ describe the annihilation of atoms and
anti-atoms with internal energy $E_n$ and center-of-mass
momentum $\hbar \vec{ k}$. They obey
the usual harmonic oscillator commutation relations
$[a_{\vec{k}}^{(n)}, a_{\vec{k}'}^{(n')}] = \delta (\vec{ k}- \vec{
  k}') \delta_{n,n'}$. We remark that for cold atoms
we have $ \lambdabar_n w_k\approx 1$, so that the particle-part 
($\propto a^{(n)}_{\vec{k}}$) of Eq.~(\ref{eq:fieldOp}) reduces to the conventional field
operator 
$\hat{\Psi}_n^{\text{(nr)}}(\vec{ x})  $ used in quantum optics.

The radiation field is described in Coulomb gauge, so that the electric
field operator contains only contributions from transverse photons,
\begin{align} 
  \vec{ E}(x) &=  \vec{ E}^{(+)}(x) +  \vec{ E}^{(-)}(x) 
\\
  \vec{ E}^{(+)}(x) &=-i \int d^3 k \sum_{\sigma=1}^2
  \sqrt{\frac{ \hbar c k_0}{2(2\pi)^3\varepsilon_0 }} 
   a_{\vec{k},\sigma} \vec{ \epsilon}_{\vec{k},\sigma} e^{-i k\cdot x},
\end{align} 
with $\vec{ E}^{(-)}(x) = \vec{ E}^{(+)\dagger }(x) $ as well as 
$k_0 =|\vec{ k}|$ and 
$[a_{\vec{k},\sigma}, a_{\vec{k}',\sigma'}] = \delta (\vec{ k}- \vec{
  k}') \delta_{\sigma\sigma'}$.  The operator $a_{\vec{k},\sigma} $
annihilates a photon with momentum $\hbar\vec{ k}$ and polarization
vector $ \vec{ \epsilon}_{\vec{k},\sigma}$.
Coulomb gauge is almost universally used
to describe the physics of atoms and light. It is not
covariant, but it does not break causality since the commutator
between transverse electric fields has support on the light cone. 

We describe the interaction between atoms and light using electric dipole
coupling in rotating-wave approximation \cite{FRASCA2003193},
\begin{align} 
  \hat{H}_\text{int}(x^0) &= \int d^3x\, \hat{H}_\text{int}(x)
\\
  \hat{H}_\text{int}(x) &= -
   \vec{ E}(x) \cdot
  \left ( \vec{d}_{eg} \hat{\Psi}_e^\dagger (x) \, \hat{\Psi}_g(x)
  + \text{h.c.} 
  \right ),
\label{eq:Hint}\end{align} 
where $x = (x^0, \vec{ x})$.
In this expression, $\vec{d}_{eg}$ denotes matrix element $\langle e |
\vec{ d} | g \rangle $ of the atomic dipole moment operator $\vec{
  d}$. In our model, it is a parameter that is fixed using experimental observations.

\begin{figure}
\begin{center}
\includegraphics[width=6cm]{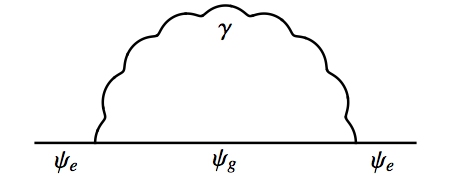}
\caption{\label{fig:T2} Self energy diagram corresponding to spontaneous emission. An
excited atom $\psi_e$ emits, and later re-absorbs, a photon $\gamma$
and turns into a ground
state atom $\psi_g$.}
\end{center}
\end{figure}

To describe spontaneous emission, we consider as initial state
a single excited atom in a vacuum. Since $\hat{H}_\text{int}$
can only excite or de-excite an atom, $\hat{T}_2$ describes a self-energy
term as depicted in Fig.~\ref{fig:T2}, in which an excited atom emits
and re-absorbs a single photon.
If we would use Eq.~(\ref{eq:retardedNaive}) to split $T_2(x,y)$ of Eq.~(\ref{eq:T2def})
in retarded and advanced part, we would have to employ renormalization
to remove divergent integrals. In CPT, $\hat{T}_2$ is instead constructed as
follows. One first notes that the following advanced and retarded distributions,
\begin{align} 
  A_2(x,y) &= A_2'(x,y)+T_2(x,y)
\\
  R_2(x,y) &= R_2'(x,y)+T_2(y,x)
\end{align} 
vanish for $ x^0>y^0$ and  $x^0<y^0$, respectively.
Forming 
\begin{align} 
  D_2(x,y) &= R_2(x,y) - A_2(x,y) = R'_2(x,y) - A'_2(x,y),
\label{eq:D2def}\end{align} 
we can find $R_2$ by taking the retarded part of $D_2$. This is done
in the next section by employing distribution splitting.

\begin{figure}
\begin{center}
\includegraphics[width=8cm]{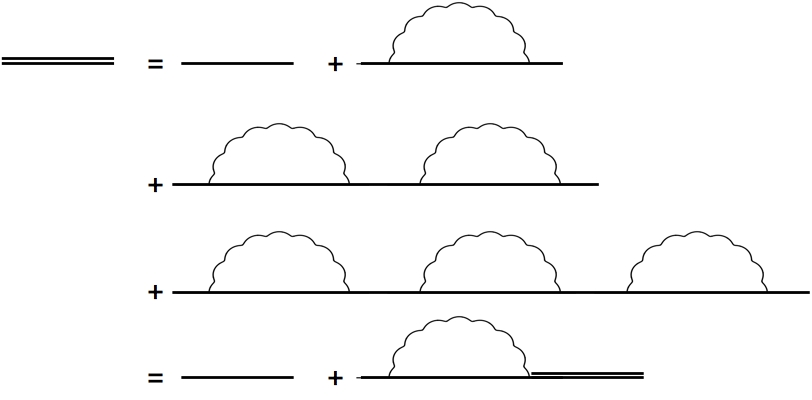}
\caption{\label{fig:dyson} Ladder approximation for the full
  propagator for excited atoms, which is represented by a double line.
  A single horizontal line denotes the unperturbed propagator. The
  last line corresponds to Dyson's equation.}
\end{center}
\end{figure}

To describe the full dynamics of a system, one would have to to
evaluate all orders of perturbation theory, which is not
feasible. However, Knight and Allen \cite{Knight197299} have shown
that the Weisskopf-Wigner theory
\cite{WignerWeisskopf1,WignerWeisskopf2} of spontaneous emission 
is equivalent to the ladder approximation in quantum field theory,
which is depicted in Fig.~\ref{fig:dyson}. It is therefore sufficient
to find $\hat{T}_2$ in order to describe spontaneous emission.

\section{Atomic decay rate and line shift}\label{sec:T2}
In CPT, it is essential that distributions are causal
in the sense that their support lies on the light cone or is
time-like. To verify that the support of an operator-valued
distribution is causal, one expresses all operator products through 
normally ordered products by using
\begin{align} 
  E_i(x) E_j(y) &= : E_i(x) E_j(y): -i D_{ij}^{(+)}(x-y) \; , \; i=1,2,3
\\
  \hat{\Psi}_i(x) \hat{\Psi}_i^\dagger (y) &=
  : \hat{\Psi}_i(x) \hat{\Psi}_i^\dagger (y) : 
  -iD_i^{(+)}(x-y) \; , \; i=e,g
\\
  \hat{\Psi}_i^\dagger (x) \hat{\Psi}_i (y) &=
  : \hat{\Psi}_i^\dagger (x) \hat{\Psi}_i (y) : 
  -iD_i^{(+)}(x-y) \; , \; i=e,g,
\end{align} 
where
\begin{align} 
   D_{ij}^{(+)} (x) &= \int d^3k \frac{\hbar c k_0
                             }{2(2\pi)^3\varepsilon_0}
   \left (\delta_{ij}  - \frac{ k_i k_j}{k^2}
  \right ) e^{-ik\cdot x}
\label{eq:DijPlus}\\
   D_i^{(+)}(x) &=\frac{2 i}{\lambdabar_i (2\pi)^3} \int d^4 k\,
  \delta (k\cdot k-\lambdabar_i^{-2}) \theta(k_0) e^{-ik\cdot x}.
\end{align} 
Applying this to $D_2(x,y)$ of Eq.~(\ref{eq:D2def}) yields
\begin{align} 
   D_2 (x,y) &=
   [ : \hat{\Psi}_e^\dagger (x) \hat{\Psi}_e (y) :
   + : \hat{\Psi}_e^\dagger (y) \hat{\Psi}_e (x) :  ] {\cal D}_2(x-y) 
\nonumber \\ & \hspace{4mm}
  + \text{rest}
\\
  {\cal D}_2(x) &=  \frac{ d_{{eg},i}d_{{eg},j}^*}{\hbar^2 c^2}
  \big ( 
    D_g^{(+)}(x)  D_{ij}(x)
  -D_g(x)  D^{(-)}_{ij}(x) 
  \big ),
\end{align} 
with $D^{(-)}(x)= - D^{(+)}(-x)$. Distributions of the form
$D = D^{(+)} + D^{(-)}$ have causal support,
so that ${\cal D}_2(x)$ is causal as well. The terms labelled ``rest''
either contain vacuum diagrams, which we
may ignore, or are proportional to normally ordered products of
ground state atoms or radiation field. Since our system initially
does not contain photons or ground state atoms, these terms 
annihilate the initial state.

To split ${\cal D}_2(x)$  into retarded and advanced parts, we have to
determine its singular order. This is best done in momentum
space. We will denote the Fourier transformation of a
function $f(x)$ by $\tilde{f}(p)$ 
\footnote{Strictly speaking, we are using wavevector $\vec{ k}$
  instead of momentum $\hbar \vec{ k}$ in all of our
  derivations.}. We find
\begin{align} 
 \tilde{ {\cal D}}_2(p) &=\frac{i(p\cdot p \lambdabar_g^{2} -1)^3
         }{12 \varepsilon_0 \hbar
           c(2\pi p\cdot p)^3  \lambdabar_g^{7} } 
  \text{sgn}(p_0)\theta(p\cdot p \lambdabar_g^{2}-1 )
\nonumber \\ &\hspace{4mm}
   \times  \left (
  |\vec{ d}_{eg}|^2 (2p_0^2-p\cdot p) -2 |\vec{ p}\cdot \vec{ d}_{eg}|^2
  \right ).
\end{align} 
For $p\rightarrow \infty$, this distribution scales as $p^\omega$
with singular order $\omega=2$.

The general method to find the retarded part $r(x)$ of a distribution $d(x)$ includes two steps. First,
$d(x)$ is multiplied with a smooth smeared-out step function
$\theta_L(x\cdot v)$, which is monotonous and takes the value 0 when the argument is negative and 1
if it is larger than $L$.  $v$ is an arbitrary time-like four
vector. For fixed $L$, the product $\theta_L(x\cdot v)d(x)$ is a well-defined distribution,
but it will diverge like $L^{-(4+\omega)}$ for $L\rightarrow
0$. Despite of this, the linear functional
\begin{align} 
   \int d^4 x \, \theta_L(x\cdot v)d(x) \, f(x)
\end{align} 
is well defined in the limit as long as all derivaties of test
function $f(x)$ up to order $\omega$ vanish at $x=0$.
To warrant that this is the case for arbitrary test functions,
$\theta_L(x\cdot v)d(x)$ is modified so that contributions of these
derivatives are removed. In momentum space, this is accomplished by
evaluating the following integral (proposition 3.4 of Ref.~\cite{ScharfQED}),
\begin{align}
  \tilde{r}(p_0) &= \frac{ i}{2\pi} p_0^{\omega+1} \int dk_0\, 
 \frac{ \tilde{d}(k_0)}{(k_0-i0)^{\omega+1}(p_0-k_0+i 0)} .
\label{eq:distSplit}\end{align} 
In this expression, the temporal axis for the $k$-integration has been
rotated by an orthogonal transformation so that it is parallel to
(time-like) four vector $p_\mu$. 

It will be explained below that for a very slow atom we can restrict
our considerations to the case $\vec{ p}=0$. To simplify the notation,
we set $p_0 = u \lambdabar_g^{-1}$ with real dimensionless parameter $u$.
We then find the for retarded part $\tilde{{\cal R}}_2(u)$ of 
$\tilde{{\cal D}}_2(p_0)$ the expression
\begin{align} 
\tilde{{\cal R}}_2(u) &=
   \frac{| d_{eg}|^2 }{6(2 \pi)^4 \hbar  c \varepsilon_0 \lambdabar_g^3}
   \Bigg[ \frac{(u^2 -1)^3 
 }{2   u^4} 
  \Big [ 
   2\pi i \theta(u^2-1)  \text{sgn} u
\nonumber \\ &\hspace{4mm}
- \log (( u^2-1)^2)\Big ]
  + \frac{1}{2  u^2}
  -\frac{5}{4} 
  + \frac{ 11}{12}u^2  
\Bigg ].
\label{eq:R2a}\end{align} 

To achieve our goal of deriving $T_2(x,y)$, we need to subtract
$R'(x,y)$ from $R_2(x,y)$. Using the same methods as for $D_2(x,y)$
we obtain
\begin{align} 
   R_2' (x,y) &=
   [ : \hat{\Psi}_e^\dagger (x) \hat{\Psi}_e (y) :
   + : \hat{\Psi}_e^\dagger (y) \hat{\Psi}_e (x) :  ] {\cal R}_2'(x-y) 
\nonumber \\ & \hspace{4mm}
  + \text{rest}
\\
  {\cal R}_2'(x) &=  -\frac{  d_{{eg},i}d_{{eg},j}^*}{\hbar^2 c^2}
   D_g^{(-)}(x)  D^{(-)}_{ij}(x) .
\end{align} 
The Fourier transform is given by
\begin{align} 
  \tilde{{\cal R}}_2'(p) &=\frac{ (p\cdot p \lambdabar_g^2 -1)^3
           }{12i \varepsilon_0 \hbar
           c(2\pi p\cdot p)^3  \lambdabar_g^7 } 
  \theta(-p_0)  \theta(p\cdot p \lambdabar_g^2-1 )
\nonumber \\ &\hspace{4mm}
   \times  \left (
  |\vec{ d}_{eg}|^2 (2p_0^2-p\cdot p) -2 |\vec{ p}\cdot \vec{ d}_{eg}|^2
  \right ).
\label{eq:FTR2p}\end{align} 
We thus arrive at
\begin{align} 
   T_2 (x,y) &=
   [ : \hat{\Psi}_e^\dagger (x) \hat{\Psi}_e (y) :
   + : \hat{\Psi}_e^\dagger (y) \hat{\Psi}_e (x) :  ] {\cal T}_2(x-y) 
\nonumber \\ & \hspace{4mm}
  + \text{rest}
\\
  {\cal T}_2(x) &=  {\cal R}_2(x) - {\cal R}_2'(x),
\end{align} 
with the Fourier transform of the right-hand side given by
Eqs.~(\ref{eq:R2a}) and (\ref{eq:FTR2p}). However, 
in appendix \ref{app:T2action} we show that 
atoms are only affected by the combination
\begin{align} 
   \tilde{{\cal T}}_2^{(\text{s})}(p) &=
  \frac{ 1}{2} ( \tilde{{\cal T}}_2(p) +  \tilde{{\cal T}}_2(-p)).
\end{align}  
For a resting atom 
with $p_\mu =(u \lambdabar_g^{-1},\vec{ 0})$, this is given by
\begin{align} 
  \tilde{{\cal T}}_2^{(\text{s})} (u) &=
   \frac{ |\vec{ d}_{eg}|^2}{12(2\pi)^4\varepsilon_0 c
  \hbar \lambdabar_g^3}
    \Bigg[ \frac{(u^2 -1)^3 
 }{2   u^4} 
  \Big [ 
   2\pi i \theta(u^2-1)  
\nonumber \\ &\hspace{4mm}
- \log (( u^2-1)^2)\Big ]
  + \frac{1}{ u^2}
  -\frac{5}{2} 
  + \frac{ 11}{6}u^2  
\nonumber \\ &\hspace{4mm}
  +C_0 +C_1 u +C_2 u^2
\Bigg ].
\label{eq:T2results}\end{align} 
Here we have added a polynomial $C_0 +C_1 u +C_2 u^2$
because distribution splitting is not unique for singular orders
$\omega \geq 0$. Different approaches to splitting a distribution
may then differ by a polynomial in $p$ of order $\omega$
\cite{ScharfQED}.
The coefficients $C_i, i=0,1,2$  play a similar role as
renormalization parameters in the usual theory of renormalization.
They are fixed by physical, rather than mathematical, considerations.

With this result, we are in the position to derive spontaneous
emission rate and line shift for the model under consideration.
It is shown in appendix \ref{app:T2action} that operator $\hat{T}_2$
acts on a slow excited atom with center-of-mass wavefunction
$\phi(\vec{ x})$ like 
$ \hat{T}_2|\phi \rangle \approx Z|\phi\rangle$,
with complex factor
\begin{align} 
Z &\approx
  2 (2\pi)^{2}  c t_g
  \tilde{{\cal T}}_2^{(\text{s})} (u),
\label{eq:T2phi}\end{align} 
where $u=\frac{ \lambdabar_g}{\lambdabar_e}$ and 
$t_g$ is the duration of the interaction between atom and radiation.
Since $m_e=m_g+\hbar \omega_{eg}/c^2$, we have
$\lambdabar_e^{-1} =\lambdabar_g^{-1}+\omega_{eg}/c$.  
We therefore can set
$u=1+\delta u$, with positive $\delta u = \hbar \omega_{eg}/(m_gc^2) \ll 1$.
The spontaneous emission rate $\gamma=\text{Im}(Z/t_g)$ is given by
\begin{align} 
  \gamma &= 
  \frac{\delta u^3 (2+\delta u)^3
   |d_{eg}|^2}{24 \pi  (1+\delta u)^5
   \epsilon _0 \hbar  \lambdabar _g^3}.
\end{align} 
To leading order in $\delta u$ we find
\begin{align} 
  \gamma &= 
  \frac{ |d_{eg}|^2 \omega_{eg}^3}{3 \pi 
   \hbar \epsilon _0 c^3}.
\end{align} 
This result agrees with the standard result for the decay rate of a
two-level atom \cite{MilonniQuantumVacuum} and does not depend on the
distribution splitting scheme.

The atomic line shift $\Delta =\text{Re}(Z/t_g)$ is given by
\begin{align} 
  \Delta &= 
  \frac{|d_{eg}|^2}{144 \pi^2  \epsilon _0 \hbar  \lambdabar _g^3}
\bigg [
  -48 \delta u^3 \log (2 \delta u)
-3 (2 C_0+15) \delta u^3
\nonumber \\ &\hspace{4mm}
  + 2 
  +6 (C_0+C_1+ C_2)
   +(8-6 C_0+6 C_2) \delta u
\nonumber \\ &\hspace{4mm}
  +3 (2C_0+7) \delta u^2
\bigg ] + {\cal O}(\delta u^4).
\label{eq:lineShift0}\end{align} 
At this point, we have to specify the normalization parameters
$C_i$. For optical transitions, parameter $\delta u$ is extremely
small ($10^{-8}-10^{-9}$), so that lower powers of $\delta u$ will
give larger contributions. We compare our result to a more precise 
expression for the Lamb shift for the 1s$\rightarrow$2p 
transition in Hydrogen \cite{0953-4075-29-2-008},
\begin{align} 
  \Delta_L &\approx \frac{ m_{\text{elec}}c^2 \alpha^5 }{\pi\hbar}
  \left (-25.25 
   + \frac{ 4}{3} \ln(\alpha^{-2})
  \right ),
\label{eq:LambShift}\end{align} 
where $\alpha \approx 1/137$ is the fine structure constant and
$m_{\text{elec}}$ the mass of the electron. It is not hard to see that,
in Eq.~(\ref{eq:lineShift0}), terms of order $\delta u^2$ or lower
will produce results that are many orders of magnitude larger than the
observed line shift. Our distribution splitting scheme is therefore fixed by
the condition that these terms disappear, which is achieved for
$C_0=-\frac{7}{2}, C_1=-\frac{ 29}{6}$, and $C_2=8$. The final result
for the line shift in a two-level model is then given by
\begin{align} 
    \Delta &= -\frac{ \gamma}{2\pi}
   \left [ 1+2 
   \log \left (
   \frac{ 2\hbar\omega_{eg}}{m_gc^2}
   \right )
   \right ].
\label{eq:lineShiftResult}\end{align} 
For a numerical comparison with a more precise prediction we use
data for the 1s$\rightarrow$2p 
transition in Hydrogen, i.e., 
$\hbar\omega_{eg}=0.75\times 13.6$ eV and 
$d_{eg} = \sqrt{2} 2^7 3^{-5}e a_0$, with $a_0$ the Bohr radius. We find
\begin{align} 
  \frac{ \Delta}{\Delta_L} &= 
 \frac{-0.029 \log (\alpha )-0.015 \log
   \left(\frac{m_e}{m_g}\right)-0.0031}{\log (\alpha
   )+9.47}
\\ &
  \approx 0.055 .
\label{eq:shiftRatio}\end{align} 
The fact that the simple two-level model makes a prediction that is
off by one order of magnitude is not surprising. The model only
includes the line shift generated by a single transition, instead of a
sum over all excited states. 
Furthermore, models like that of Eq.~(\ref{eq:QOexcHam}),
where excitation and de-excitation are treated separately, can be
considered as a consequence of the rotating-wave approximation, which
neglects further off-resonant contributions to the line shift.

\section{Discussion}\label{sec:discussion}
In the language of renormalization and regularization, one would identify the
atomic Compton angular frequency
$m_g c^2/\hbar$, which appears in the logarithm in
Eq.~(\ref{eq:lineShiftResult}), with a cutoff frequency.
CPT therefore predicts that in models where atoms are
  considered as point dipoles, it is the atomic mass that determines
  the cutoff. However, in most applications of atomic point-dipole models,
  the line shift is either ignored, or the cutoff is taken to be the
  Compton frequency of the electron, so that the results resemble the
  Lamb shift (\ref{eq:LambShift}). It appears that, for atomic
  point-dipole models to be consistent with causality, the
  atomic Compton frequency should be used instead.

Another point of interest about our findings is connected to the
choice (\ref{eq:Hint}) of the interaction between atoms and
radiation. Electric dipole coupling $- \vec{ d}\cdot \vec{ E}$ is
popular in quantum optics, but in relativistic theories minimal
coupling ($\vec{ p}\cdot \vec{ A}$, with $\vec{ p}$ the momentum and
$\vec{ A}$ the vector potential) is preferred. 
However, using minimal coupling would only be possible in Lorentz
gauge, which is rarely used to describe non-relativistic atoms. 
In Coulomb gauge, the vector
potential will be a transverse field, i.e., it obeys $\nabla\cdot
\vec{ A}=0$. It can be shown that the commutator between transverse
vector potentials has space-like support \cite{PhysRevA.58.3407}, so
that CPT cannot 
be used for minimal coupling in Coulomb gauge.

One of the motivations for this work was the question whether
separating a vector field operator $\vec{ V}(\vec{ r})$
into a transverse part $\vec{ V}_\perp$ and a longitudinal part
$\vec{ V}_\| = \vec{ V}-\vec{ V}_\perp$ would require the methods of
distribution splitting.
The transverse part
is most easily constructed in momentum space by modifying
its Fourier coefficients as 
$ \tilde{V}_{\perp,i}(\vec{ k}) = (\delta_{ij} -k_ik_j/k^2)\tilde{V}_{j}(\vec{ k})$,
similar to Eq.~(\ref{eq:DijPlus}). This modification does not change
the singular order of a distribution, so that the separation into
transverse and longitudinal parts can be performed in a conventional way.

Our last remark concerns the question whether the model discussed in
this work is renormalizable, i.e., whether a finite set of
normalization parameters is sufficient to describe the S-matrix to all
orders in perturbation theory. While the answer to this question is
beyond the scope of this work, we can offer the following comments.

Renormalizability hinges on the presence of symmetries in a theory and
is usually proven by exploiting 
Ward-Takahashi identities \cite{dutsch2004}. The model
we have studied is gauge invariant since the interaction Hamiltonian
couples to the electric field. Charge is trivially conserved since all
atoms are electrically neutral. However, the lack of Lorentz
invariance may reduce the symmetry of our theory in such a way that
renormalizability is not warranted anymore.

\section{Conclusion}
We have applied causal perturbation theory to
derive spontaneous emission rate and line shift for a two-level atom
coupled to the radiation field. With this method, the appearance of
divergent integrals is avoided by employing the mathematical technique
of distribution splitting. Our result for the atomic decay rate agrees
with the results based on ordinary perturbation theory. This is to be
expected since the decay rate does not depend on the normalization
procedure in CPT, or renormalization parameters in other approaches.

The result for the line shift is comparable to other predictions in its
structure. However, it suggests that the cutoff frequency is
determined by the atomic mass instead of electron mass or Bohr radius.
This difference arises because, in our effective theory, atoms are
treated as point dipoles with inner structure, so that it is the mass
of this object that is relevant.

The problem we considered serves as an illustration how CPT can be
applied to effective models in quantum optics or condensed matter theory. Its
rigorous approach may be useful to study models in which other methods
produce contradicting results. We are planning to apply CPT to
other systems in quantum optics, such as dense atomic gases or atoms
in absorbing dielectrics.

\acknowledgements
We are grateful to Mark Walton for discussions about this work, and to
the Natural Sciences and Engineering Research
Council of Canada (NSERC) for financial support. B.~F.~thanks Saint
Francis Xavier University for a UCR summer research award.

\appendix
\section{The action of $\hat{T}_2$ on a slow atom}\label{app:T2action}
$\hat{T}_2$ has the general form
\begin{align} 
  \hat{T}_2 &= \int d^4 x\, d^4 y\,
  {\cal T}_2(x-y) g(x)\, g(y)
\nonumber \\ &\times \left (
  : \hat{\Psi}_e^\dagger (x) \hat{\Psi}_e(y):
  +  : \hat{\Psi}_e^\dagger (y) \hat{\Psi}_e(x):
  \right ),
\end{align} 
and is applied to a state that describes a single
non-relativistic excited atom,
\begin{align} 
  |\phi \rangle &= \int d^3k\, 
   \tilde{\phi}(\vec{ k}) a_{\vec{k}}^{(e)\dagger} |0 \rangle .
\end{align} 
Using Eq.~(\ref{eq:fieldOp}) we obtain
\begin{align} 
  \hat{\Psi}_e(x)  |\phi \rangle &=
   \phi(x) |0 \rangle 
\\
  \phi(x) &=
   (2\pi)^{-\frac{ 3}{2}} \int \frac{ d^3 k}{\sqrt{\lambdabar_ew_k}}
                                   e^{-ik\cdot x} 
  \tilde{\phi}(\vec{k}) ,
\end{align} 
where the time component of four vector $k$ is given by
  $w_k$ (this convention is throughout this appendix).
This leads to
\begin{align} 
  \hat{T}_2|\phi \rangle &=
   \int d^4 x\, d^4 y\,
  ({\cal T}_2(x-y)+{\cal T}_2(y-x))
\nonumber \\ &\hspace{4mm}\times
   \phi(y) g(y) 
    g(x) \hat{\Psi}_e^\dagger (x) 
 |0 \rangle .
\end{align} 
In momentum space, this expression takes the form
\begin{align} 
 \hat{T}_2|\phi \rangle &=
   \int \frac{ d^4p}{(2\pi)^2} 
   (\tilde{{\cal T}}_2(p)+ \tilde{{\cal T}}_2(-p))
   \int d^4 y\, e^{ip\cdot y} g(y)\phi(y)
\nonumber \\ &\hspace{4mm}\times
  \int d^4 x\, e^{-ip\cdot x} g(x) \hat{\Psi}_e^\dagger (x) 
 |0 \rangle 
\\ &=
     \int \frac{ d^4p}{2\pi} 
   (\tilde{{\cal T}}_2(p)+ \tilde{{\cal T}}_2(-p))
  \int \frac{  d^3k}{\sqrt{\lambdabar_e w_k}}
  \tilde{\phi}(\vec{ k}) \tilde{g}(p-k)
\nonumber \\ &\hspace{4mm}\times
  \int \frac{  d^3k'}{\sqrt{\lambdabar_e w_{k'}}}
  \tilde{g}(k'-p) a_{\vec{ k}'}^{(e)\dagger} |0 \rangle 
\end{align} 
We now assume that $g(x) $ does not vary much over
the spatial support of $\phi(\vec{ x})$, so that we can set  
$\phi(\vec{ x}) g(x) \approx \phi(\vec{ x}) g(x^0)$.
Using the convolution theorem, this relation implies that, in momentum
space,
\begin{align} 
  \tilde{\phi}(\vec{ k}) \tilde{g}(p-k)
  \approx (2\pi)^{3/2} 
   \tilde{\phi}(\vec{ k})
   \delta(\vec{ p}-\vec{ k}) 
  \tilde{g}(p_0-k_0).
\end{align} 
Hence,
\begin{align} 
 \hat{T}_2|\phi \rangle &\approx
 \sqrt{2\pi}\int  d^4p\,
   (\tilde{{\cal T}}_2(p)+ \tilde{{\cal T}}_2(-p))
  \frac{\tilde{\phi}(\vec{ p}) }{\sqrt{\lambdabar_e  w_p}}
\tilde{g}(p_0-w_p)
\nonumber \\ &\hspace{4mm}\times
  \int \frac{  d^3k'}{\sqrt{\lambdabar_e  w_{k'}}}
  \tilde{g}(k'-p) a_{\vec{ k}'}^{(e)\dagger} |0 \rangle 
\\ &\approx
   (2\pi)^2\int  \frac{ d^4p}{\lambdabar_e w_p}
   (\tilde{{\cal T}}_2(p)+ \tilde{{\cal T}}_2(-p))
  \tilde{\phi}(\vec{ p}) 
\tilde{g}(p_0-w_p)
\nonumber \\ &\hspace{4mm}\times
  \tilde{g}(w_p-p_0)
   a_{\vec{p}}^{(e)\dagger} |0 \rangle .
\end{align} 
If the width of wavepacket $ \tilde{\phi}(\vec{ k})$ in momentum
space is so narrow that dispersion can be neglected, 
we can set $w_p=\sqrt{\vec{ p}^2+\lambdabar_e^{-2}}\approx
\lambdabar_e^{-1}$. This approximation is well justified
  for typical atomic gases with temperatures at or below room temperature.
We furthermore assume that over the width of $ \tilde{\phi}(\vec{ k})$
we can neglect the dependence of $\tilde{{\cal T}}_2(p)$ on the
spatial components of $p$, so that
$\tilde{{\cal T}}_2(p)\approx \tilde{{\cal T}}_2(p_0)$. The action of
$\hat{T}_2$ then simplifies to
\begin{align} 
 \hat{T}_2|\phi \rangle &\approx
   (2\pi)^2 
   \int   dp_0\,
   (\tilde{{\cal T}}_2(p_0)+ \tilde{{\cal T}}_2(-p_0))
\nonumber \\ &\hspace{4mm}\times
\tilde{g}(p_0-\lambdabar_e^{-1})
  \tilde{g}(\lambdabar_e^{-1}-p_0)
   \int  d^3p\,
  \tilde{\phi}(\vec{ p}) 
   a_{\vec{p}}^{(e)\dagger} |0 \rangle 
\\ &= Z |\phi \rangle 
\\
  Z &=  (2\pi)^2 
   \int   dp_0\,
   (\tilde{{\cal T}}_2(p_0)+ \tilde{{\cal T}}_2(-p_0))
\nonumber \\ &\hspace{4mm}\times
\tilde{g}(p_0-\lambdabar_e^{-1})
  \tilde{g}(\lambdabar_e^{-1}-p_0).
\end{align} 
If $\tilde{g}(p_0-\lambdabar_e^{-1})$ is sufficiently narrow, we can
set $\tilde{{\cal T}}_2(p_0)\approx \tilde{{\cal
    T}}_2(\lambdabar_e^{-1})$.
With this approximation we obtain
\begin{align} 
  Z&\approx  
   (2\pi)^2 
   (\tilde{{\cal T}}_2(\lambdabar_e^{-1})+ \tilde{{\cal T}}_2(-\lambdabar_e^{-1}))
\nonumber \\ &\hspace{4mm}\times
     \int_{-\infty}^\infty   dp_0\,
   \tilde{g}(p_0-\lambdabar_e^{-1})
  \tilde{g}(\lambdabar_e^{-1}-p_0)
\\ &=
    (2\pi)^2 
   (\tilde{{\cal T}}_2(\lambdabar_e^{-1})+ \tilde{{\cal
     T}}_2(-\lambdabar_e^{-1}))
   \int_{-\infty}^\infty dx^0\, g^2(x^0).
\end{align} 
If test function $g(x^0)$ is close to unity during a time interval of width
$c t_g$ and drops quickly to zero outside the interval, the integral
in this expression is approximately equal to $c t_g$.
Using this approximation results in Eq.~(\ref{eq:T2phi}).

\bibliographystyle{apsrev4-1}
\bibliography{NRQED8.bib}
\end{document}